\documentclass[nofootinbib,amsfonts,prd,aps]{revtex4}
\usepackage{graphicx}
\begin{document}

\title{Quantum space-time of a charged black hole}

\author{Rodolfo Gambini$^{1}$, Esteban Mato Capurro$^{1}$, 
Jorge Pullin$^{2}$}
\affiliation {
1. Instituto de F\'{\i}sica, Facultad de Ciencias, 
Igu\'a 4225, esq. Mataojo, 11400 Montevideo, Uruguay. \\
2. Department of Physics and Astronomy, Louisiana State University,
Baton Rouge, LA 70803-4001}

\begin{abstract}
  We quantize spherically symmetric electrovacuum gravity.  The
  algebra of Hamiltonian constraints can be made Abelian via a
  rescaling and linear combination with the diffeomorphism
  constraint. As a result the constraint algebra is a true Lie
  algebra. We complete the Dirac quantization procedure using loop
  quantum gravity techniques. We present explicitly the exact
  solutions of the physical Hilbert space annihilated by all
  constraints.  The resulting quantum space-times resolve the
  singularity present in the classical theory inside charged black
  holes and allows to extend the space-time through where the
  singularity used to be into new regions. We show that quantum
  discreteness of space-time may also play a role in stabilizing the
  Cauchy horizons, though back reaction calculations are needed to
  confirm this point.
\end{abstract}

\maketitle
\section{Introduction}

Charged black holes are not expected to play a significant role in
astrophysics, but they are a good laboratory to test important ideas
in black hole physics. Unlike neutral Schwarzschild black holes,
charged Reissner-Nordstrom black holes share elements in common with
rotating black holes, like the appearance of Cauchy horizons. Vacuum
Schwarzschild black holes have been recently treated using loop
quantum gravity techniques \cite{sphericalprl}. Key to being able to
quantize these systems was the realization that one can linearly
combine the Hamiltonian and diffeomorphism constraints into
constraints that satisfy a Lie algebra. This allows the completion of
the Dirac quantization program. Perhaps more surprising, the physical
space of states was found in closed form. New observables that do not
have a classical counterpart appear in the quantum theory. The metric
of space-time can be written as an operator associated with a
parametrized Dirac observable acting on the space of physical
states. Analyzing the metric, it was found that the singularity is
resolved by quantum effects and one tunnels into another region of
space-time through a region where the singularity used to be in the
classical theory where quantum effects are not negligible.

The purpose of this paper is to show that the above results can be
extended to the case of charged spherically symmetric black holes. We
will see that the singularity is again resolved by the quantum
theory. In addition to that, new perspectives on the stability of
Cauchy horizons arise.

\section{Spherically symmetric electrovac gravity: the classical theory}

The treatment of spherically symmetric space-times with Ashtekar-type
variables was pioneered by Bengtsson \cite{bengtsson} and in more
modern language discussed in detail by Bojowald and
Swiderski \cite{boswi}. We will follow here the notation of our
previous paper \cite{spherical} and we refer the reader to them and to
Bojowald and Swiderski for more details.

Ashtekar-like variables adapted to the symmetry of the problem, after
some work, lead to two pairs of canonical variables $E^\varphi$,
${K}_\varphi$ and $E^x$, $K_x$, that are related to the
traditional canonical variables in spherical symmetry $ds^2=\Lambda^2
dx^2+R^2 d\Omega^2$ by $\Lambda=E^\varphi/\sqrt{|E^x|}$, $P_\Lambda=
-\sqrt{|E^x|}K_\varphi$, $R=\sqrt{|E^x|}$ and $P_R=-2\sqrt{|E^x|} K_x
-E^\varphi K_\varphi/\sqrt{|E^x|}$ where $P_\Lambda, P_R$ are the
momenta canonically conjugate to $\Lambda$ and $R$ respectively, $x$
is the radial coordinate and $d\Omega^2=d\theta^2+\sin^2\theta
d\varphi^2$. We consider a spherically symmetric electromagnetic field
${\bf A}= \Gamma dr+\Phi dt$ paramterized by two configuration
variables $\Gamma,\Phi$ and their canonically conjugate momenta,
$P_\Gamma,P_\Phi$. We assume a trivial bundle for the electromagnetic
field implying the absence of monopoles. In the canonical treatment it
is found that $\Phi$ operates as a Lagrange multiplier, and can be
dropped as a canonical variable \cite{loukowintershilt}.

The constraints of the theory are given by the Hamiltonian,
diffeomorphism and electromagnetic Gauss law constraints,
\begin{eqnarray}
  H &=& 
-\frac{E^\varphi}{2\sqrt{E^x}} - 2 K_\varphi \sqrt{E^x} K_x  
-\frac{E^\varphi K_\varphi^2}{2 \sqrt{E^x}}+\frac{\left((E^x)'\right)^2}{8\sqrt{E^x}E^\varphi}
-\frac{\sqrt{E^x}(E^x)' (E^\varphi)'}{2 (E^\varphi)^2} +
\frac{\sqrt{E^x} (E^x)''}{2 E^\varphi} +G \frac{E^\varphi}{2
  \left(E^x\right)^{3/2}} P_\Gamma^2,\\
 C &=& -(E^x)' K_x +E^\varphi (K_\varphi)'-G\, \Gamma P_\Gamma',\\
 {\cal G} &=& P_\Gamma',
\end{eqnarray}
where we have chosen the Immirzi parameter to one. 
We proceed to rescale the Lagrange multipliers,
$  N_r^{\rm old}=N_r^{\rm new} -2 N^{\rm
  old}\frac{K_\varphi\sqrt{E^x}}{\left(E^x\right)'}$ and 
  $N^{\rm old} = N^{\rm new} \frac{\left(E^x\right)'}{E^\varphi}$,
and from now onwards we will drop the ``new'' subscripts for brevity. 
This leads to a
total Hamiltonian,
\begin{eqnarray}
H_T &&=\int dx \left\{ -N
\left[\left(-\sqrt{E^x}\left(1+K_\varphi^2\right)+\frac{\left(\left(E^x\right)'\right)^2\sqrt{E^x}}{4
    \left(E^\varphi\right)^2}+2 G M\right)'+G \frac{\left(E^x\right)'}{2
  \left(E^x\right)^{3/2}}P_\Gamma^2+2 G
\frac{K_\varphi}{E^\varphi}\Gamma P_\Gamma'\right]\right.\nonumber\\
&&\left.\begin{array}{c}\\ \\ \\ \end{array}+ N_r \left[-
(E^x)' K_x +E^\varphi (K_\varphi)'-\Gamma P_\Gamma'\right]
+\lambda'\left( P_\Gamma+Q\right)\right\}, 
\end{eqnarray}
with the Lagrange multipliers $N$, the lapse, $N_r$ the shift and
$\lambda$ the parameter of Gauss law.  The $GM$ and $Q$ terms are
constants of integration that arise from an examination of the theory
at spatial infinity. This is standard so we refer the reader to
previous papers for it \cite{saeed,kieferlouko}.  The rescaling makes the
Hamiltonian constraint have an Abelian algebra with itself, and the
usual algebra with the diffeomorphism constraint and Gauss law. We had
already noted this in vacuum \cite{sphericalprl}, here we point out
that it also holds with the inclusion of an electromagnetic field.

We are interested in partially fixing the electromagnetic gauge to
$\Gamma=0$, which is natural for static situations. This determines
the Lagrange multiplier $\lambda$ and also turns the Gauss law into a
strong constraint $P_\Gamma=-Q$. This leads to a 
total Hamiltonian of the form,
\begin{equation}
H_T =\int dx \left\{ -N
\left(-\sqrt{E^x}\left(1+K_\varphi^2 +\frac{G Q^2}{E^x}\right)
+\frac{\left(\left(E^x\right)'\right)^2\sqrt{E^x}}
{4 \left(E^\varphi\right)^2}+2 G M
\right)'
+ N_r \left[-
(E^x)' K_x +E^\varphi (K_\varphi)'\right]
\right\}, 
\end{equation}
where we identify the contribution of the electromagnetic field to the
mass function, proportional to $Q^2$.

Notice that if one were to choose the gauge $E^x=x^2$ and
$K_\varphi=0$ the preservation of the gauge conditions requires that
$N_r=0$ and one would get the Reissner-Nordstrom metric in
Schwarzschild form,
\begin{equation}
  ds^2=-\left(1-\frac{2 G M}{x}+\frac{G Q^2}{x^2}\right)dt^2+
  \frac{1}{1-\frac{2 G M}{x} + \frac{G Q^2}{x^2}}dr^2+x^2d\Omega^2.
\end{equation}
\section{Quantization: kinematics}

We now proceed to quantize. We start by recalling   the basis of
spin network states in one dimension (see \cite{spherical} for
details). One has  graphs $g$ consisting of a collection of edges
$e_j$ connecting
the vertices $v_j$. 
It is natural to associate the variable $K_x$ with
edges in the graph and the variable $K_\varphi$ with vertices of the
graph. For bookkeeping purposes we will associate each edge with the
vertex to its left. One then constructs the 
``holonomies'' (only $K_x$ is a true connections, so the
``holonomies'' associated with $K_\varphi$ are ``point'' holonomies),
\begin{eqnarray}
T_{g,\vec{k},\vec{\mu}}(K_x,K_\varphi) &=& \langle K_x,K_\varphi
\left\vert\vphantom{\frac{1}{1}}\right.\left.
\raisebox{-7mm}{\includegraphics[height=2.0cm]{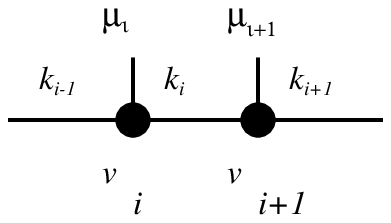}}\right\rangle\nonumber\\
&=&
\prod_{e_j\in g}
\exp\left(\frac{i}{2} k_{j} \int_{e_j} K_x(x)dx\right)
\prod_{v_j\in g}
\exp\left(\frac{i}{2}  \mu_{j}\gamma K_\varphi(v_j) \right)
\end{eqnarray}
with $e_j$ the edges of the spin network $g$ and $v_j$ its vertices
and the integer $k_j$ is the (integer) valence associated with the
edge $e_j$ and the integer number $\mu_j$ the ``valence'' associated with
the vertex $v_j$. Notice that since we gauge fixed the electromagnetic
field, the kinematical states are the same as those for vacuum
gravity.

On these states the triads act multiplicatively,
\begin{eqnarray}
{\hat{E}^x(x) } T_{g,\vec{k},\vec{\mu}}
&=& \ell_{\rm Planck}^2 k_i(x)
 T_{g,\vec{k},\vec{\mu}}
\\
\int_I\hat{E}^\varphi(x) T_{g,\vec{k},\vec{\mu}}
&=& \frac{\gamma \ell_{\rm Planck}^2}{4\pi} \sum_{v_j\in I} \mu_j
T_{g,\vec{k},
\vec{\mu}},
\end{eqnarray}
where $I$ is an interval, and $k_i(x)$ is the valence of the edge that
contains the point $x$.

The problem has two global variables, the mass and the charge. Each of
them is associated with a Hilbert space of square integrable
functions. So the complete kinematical Hilbert space is given by,
functions, the kinematical Hilbert space is given by,
\begin{equation}
  H_{\rm kin}= H^M_{\rm kin}\otimes H^Q_{\rm kin} \left[\otimes_{j=1}^V l^2_j\otimes
    l_j^2\right]
\end{equation}
with $l^2_j$ the space of square integrable functions associated with
the vertex $v_j$ and $V$ the number of vertices and $H^M_{\rm kin}$
and $H^Q_{\rm kin}$ are the Hilbert spaces associated with the mass
and charge. We have chosen periodic functions in $K_\varphi$ with
period $\pi/\rho$ with $\rho$ a real constant. As discussed in
\cite{olmedo} an equivalent quantization can be constructed choosing a
Bohr compactification. Notice that we are working with a fixed number
of vertices. This will be justified later on by noticing that the
diffeomorphism and Hamiltonian constraints do not change the number of
vertices. 

The Hilbert space is endowed with an inner product,
\begin{equation}
\langle  g,\vec{k},\vec{\mu},q, M\vert
g',\vec{k}',\vec{\mu}',q',M'\rangle
=\delta_{\vec{k},\vec{k}'}\delta_{\vec{\mu},\vec{\mu}'}\delta_{g,g'}\delta(M-M')
\delta(Q-Q')
\end{equation}
where we are not assuming the charge to be quantized.

On this space the kinematical momentum operators are multiplicative,
\begin{eqnarray}
  \hat{M} \vert g,\vec{k},\vec{\mu},Q, M\rangle&=&
M\vert g,\vec{k},\vec{\mu},Q, M\rangle,\\
  \hat{Q} \vert g,\vec{k},\vec{\mu},Q, M\rangle&=&
Q\vert g,\vec{k},\vec{\mu},Q, M\rangle,\\
\hat{E}^x(x) \vert g,\vec{k},\vec{\mu},Q, M\rangle&=&
\ell_{\rm Planck}^2 k_j(x)\vert g,\vec{k},\vec{\mu},Q, M\rangle,\\
\int_{I} dx \hat{E}^\varphi(x) \vert g,\vec{k},\vec{\mu},Q, M\rangle&=&
\sum_{v_j\in I} \ell_{\rm Planck}^2 
\mu_j \vert g,\vec{k},\vec{\mu},Q, M\rangle,
\end{eqnarray}
and the holonomies act as,
\begin{eqnarray}
  \exp\left(\frac{i n}{2} \int_{e_j} dx K_x(x)\right)
\vert g,\vec{k},\vec{\mu},Q, M\rangle &=&
\vert g,k_1,\ldots,k_j+n,\ldots,k_V,\vec{\mu},Q, M\rangle, \\
\exp\left(\pm \frac{in}{2} \rho K_\varphi(v_j)\right)
\vert g,\vec{k},\vec{\mu},Q, M\rangle &=&
\vert g,\vec{k},\mu_1,\ldots,\mu_j\pm n,\ldots,\mu_V,Q, M\rangle.
\end{eqnarray}
We are restricting the action of the holonomy of $K_\varphi$ to
vertices since acting elsewhere it would create a new vertex and we
are only interested in situations with a fixed number of vertices.

\section{Quantization: dynamics}

To deal with the Hamiltonian constraint one needs to polymerize it and
choose a factor ordering. We start with the classical expression and
integrate by parts,
\begin{equation}
  H(N)=\int dx
  N'\left[\sqrt{E^x}\left(1+K_\varphi^2+\frac{G Q^2}{E^x}\right)-2 G M -
\frac{\left(\left(E^x\right)'\right)^2\sqrt{E^x}}{4
  \left(E^\varphi\right)^2}\right].
\end{equation}
This expression can be factorized,
\begin{equation}
  H(N)= \int dx N' H_+ H_-,
\end{equation}
with 
\begin{equation}
  H_{\pm} =\sqrt{\sqrt{E^x}\left(1 +K_\varphi^2
      +\frac{G Q^2}{E^x}\right)-2G M }
\pm \frac{\left(E^x\right)'\left(E^x\right)^{1/4}}{2 E^\varphi}.
\end{equation}
We now absorb one of the two factors into the lapse and have and
rescaling by a factor of $4 \left(E^\varphi\right)^2$,
\begin{equation}
H(\bar{N})=\int dx \bar{N}\left(2 E^\varphi
  \sqrt{\sqrt{E^x}\left(1+K_\varphi^2+\frac{GQ^2}{E^x}\right)-2 G M}-{\left(E^x\right)'\left(E^x\right)^{1/4}}\right).  
\end{equation}

This expression is readily quantized choosing a factor ordering,
\begin{equation}
\hat{H}(\bar{N}) \vert \Psi_g\rangle =
\int dx \bar{N}\left(2 
  \left[\sqrt{\widehat{\sqrt{E^x}}\left(1+\frac{\widehat{\sin^2\left(\rho
            K_\varphi\right)}}{\rho^2}+\frac{GQ^2}{\hat{E^x}}\right)-2 G
      M}\right]
\hat{E}^\varphi
-{\widehat{\left(E^x\right)'}\widehat{\left(E^x\right)^{1/4}}}\right)  \vert \Psi_g\rangle. 
\end{equation}
The term involving a sine, although readily realizable, implies a
finite translation in $\vec{\mu}$ leading to an equation in finite
differences, that is not easy to solve. 
It turns out that it is much more convenient 
to study the action of the Hamiltonian constraint in a mixed
representation, where we use the connection representation in
$K_\varphi$ and the loop representation in $K_x$,
\begin{equation}
  \vert \Psi_g \rangle =\int_0^\infty dM \int_{-\infty}^\infty dQ
  \prod_{v_j\in g} \int_0^{\pi/\rho} dK_\varphi(v_j) \sum_{\vec{k}}
  \vert g, \vec{k},\vec{K}_\varphi, M, Q\rangle \psi\left(M,Q,\vec{k},\vec{K}_\varphi\right),
\end{equation}
where $\vec{K}_\varphi$ is a vector that has as i-th component
$K_{\varphi}(v_i)$. On these states $\hat{E}^\varphi=-i \ell_{\rm
  Planck}^2 \partial/\partial K_\varphi$.

We will assume that the function $\psi$ is factorizable, i.e.,
\begin{equation}
\psi\left(M,Q,\vec{k},\vec{K}_\varphi\right)=
\prod_j\psi_j\left(M,Q,k_j,k_{j-1},K_\varphi(v_j)\right). 
 \end{equation}
This does not imply loss of generality as the operator has the form of
a sum of operators each acting non-trivially only on a given
vertex.
\begin{equation}
  4 i \ell_{\rm Planck}^2 \frac{\sqrt{1+m_j^2 \sin^2 \left(y_j\right)}}{m_j}
\partial_{y_j} \psi_j 
    +\ell_{\rm Planck^2} \left(k_j-k_{j-1}\right)\psi_j=0.
\end{equation}
where $y_j=\rho K_\varphi(v_j)$ and 
\begin{equation}
  m_j^2=\rho^{-2}\left(1-\frac{2GM}{\sqrt{\ell_{\rm Planck}^2
        k_j}}+\frac{G Q^2}{\ell_{\rm Planck}^2k_j}\right).
\end{equation}

This equation can be readily solved,
\begin{equation}
  \psi_j\left(M,Q,k_j,k_{j-1},K_\varphi\left(v_j\right)\right)=
\exp\left(\frac{i}{4} m_j \left(k_j-k_{j-1}\right)F\left(\rho
  K_\varphi\left(v_j\right),i m_j\right)\right),
\end{equation}
with $F$ a function of two variables given by,
\begin{equation}
  F\left(\phi,K\right)=\int_0^\phi \frac{dt}{\sqrt{1+K^2 \sin^2 t}},
\end{equation}
with $m_j$ complex inside the black hole between the horizons. The
states are normalizable with respect to the kinematical inner
product. For a lengthier discussion of normalizability, we refer the
reader to \cite{olmedo}.

\section{Observables}

There are several immediately identified Dirac observables. 
To begin with one has the mass and charge, which are observables both at a
classical and quantum level. But in addition to them one has
observables that do not have a simple classical counterpart. The first
such observable is the number of vertices. The implementation of the
Hamiltonian constraint we chose does not change the number of vertices
when acting on states of the kinematical Hilbert space. The states of
the physical space of states, annihilated by the constraint, can be
chosen all with the same number of vertices. 

An additional observable can be hinted from the fact that
(non-singular) diffeomorphisms in one dimension will not alter the
order of the vertices. Therefore the tower of values of $\vec{k}$ is
diffeomorphism invariant and unchanged by the Hamiltonian
constraint. Therefore one can readily construct an observable
associated with this property. Consider a parameter $z$ in the
interval $[0,1]$. We define,
\begin{equation}
  \hat{O}(z) \vert \Psi\rangle_{\rm phys}=\ell_{\rm Planck}^2 k_{{\rm
      Int}(V z)} \vert \Psi\rangle_{\rm phys},
\end{equation}
where ${\rm Int}(V z)$ is the integer part of the product of $z$ times
the number of vertices. As $z$ sweeps from zero to one, it will
produce as a result the components of $\vec{k}$ in an ordered way. This
observable may sound artificial, but it actually can be used to
capture the gauge invariant portion of $E^x$. The latter is not
diffeomorphism invariant. However, if we consider a function of the
real line into the $[0,1]$ interval $z(x)$ we can define,
\begin{equation}
  \hat{E}^x(x) \vert \Psi\rangle_{\rm phys} =\hat{O}\left(z(x)\right) \vert \Psi\rangle_{\rm phys}.
\end{equation}
The result is a parametrized Dirac observable (or ``evolving constant
of the motion''). It is a Dirac observable, but its value is only well
defined if one specifies a (functional) parameter $z(x)$. Specifying
the parameter is tantamount to fixing the gauge (diffeomorphisms) in
the radial direction. This is a known mechanism \cite{hawkingus} 
for representing gauge
dependent quantities on the space of physical states, where only Dirac
observables are well defined naturally. 

Defining $\hat{E}^x$ on the space of physical states has interesting
physical quantities as it allows us to define the metric as an
operator on such space. Classically its components are given by,
\begin{eqnarray}
  g_{tx} &=& - \frac{K_{\rm \varphi} \left(E^x\right)'}
{2 \sqrt{E^x} \sqrt{\left(1+K_\varphi^2\right) -\frac{2 G M}{\sqrt{E^x}} +
    \frac{G Q^2}{E^x}}},\\
g_{xx} &=& 
 \frac{\left(\left(E^x\right)'\right)^2}
{4 E^x \left(\left(1+K_\varphi^2\right) -\frac{2 G M}{\sqrt{E^x}} +
    \frac{G Q^2}{E^x}\right)},\\
g_{tt} &=& -\left(1-\frac{2 G M}{\sqrt{E^x}}+\frac{G Q^2}{E^x}\right).
\end{eqnarray}

These expressions can be readily promoted to (parametrized) Dirac
observables acting on the space of physical states. One replaces
$E^x\to \hat{E}^x$, $M\to\hat{M}$ and $Q\to \hat{Q}$. The quantity
$K_\varphi$ remains classical, it is a (functional) parameter on which
the observable depends (it also depends on $z(x)$ through
$\hat{E}^x$). The parameter $K_\varphi$ is associated with the
slicing. This can be directly seen in $g_{tx}$. A choice $K_\varphi=0$
yields $g_{tx}=0$, that is, a manifestly static slicing. With nonzero
$K_\varphi$ one can accommodate slicings that are horizon penetrating
like Painlev\'e--Gullstrand or Kerr--Schild. 

One wishes the metric to be a self-adjoint operator. Given the square
root this would be violated if one allowed a component of $\vec{k}$ to
vanish. Fortunately, since the action of the constraints does not
connect states with vanishing values of components of $\vec{k}$ with
other states, that means we can simply exclude such states and the
operators remain well defined and are self-adjoint. Remarkably, this
implies that $r=0$ is excluded from the treatment, therefore removing
the singularity. This is similar to what we observed in vacuum. One
can then consider extending the geometry to negative values of $x$,
continuing it through the region where the singularity used to be into
a new region of space-time. The resulting Penrose diagram is similar
to the one obtained by analytic extensions \cite{stoica}.

\section{Cauchy horizons and discrete space-time}

Recalling that $E^x=R^2$, with $R$ the radius of the spheres of
symmetry, the fact that the eigenvalues of $\hat{E}^x$ are discrete
imposes a constraint on the minimum increment in the value of $R$ as
one goes from a vertex of the spin network to the next, equal to
$\ell_{\rm Planck}^2/(2 R)$. That means that in the exterior of a
black hole the maximum spacing one can have occurs close to the
horizon and is given by $\ell_{\rm Planck}^2/(4 G M)$. This
fundamental level of discreteness has implications when one studies
the propagation of waves on the quantum space-time. It implies that
transplanckian modes of very high frequencies are eliminated. The
finest lattice one can have, determined by the spin network and the
condition of the quantization of $E^x$, will be a non-uniform lattice
that gets progressively coarser towards the horizon. However,
propagation of waves on non-uniform lattices involves a series of
phenomena, like attenuation and reflection of waves. If one studies
the propagation of waves on a black hole geometry in the exterior of
the black hole, the natural coordinate to use is the tortoise
coordinate $r=2GM +\ln(r/(2 G M) -1)$, since in such coordinate one is
left with a wave equation with a potential that can be readily
analyzed. In such coordinates, the condition for the quantization of
the areas implies that the lattice points get progressively more and
more separated as one approaches the horizon \cite{hawkingus}. So the
propagation of wavepackets gets more and more disrupted as one
approaches the horizon, exhibiting attenuation and reflection. In
ordinary radial coordinates this can also be seen, there it would be
the by-product of the progressive blueshifting of the incoming modes.

This non standard propagation due to the quantum space-time may have
implications for the stability of the Cauchy horizons present in the
interior of Reissner-Nordstrom black hole
\cite{instability}. The heuristic argument for
instability of such horizons is as follows. Suppose one has two
observers in the exterior and one of them decides to enter the black
hole. The external observer remains static and shines a flashlight on
the infalling observer. By the time the infalling observer reaches the
inner Cauchy horizon, the observer in the outside reaches $i^+$. That
means the exterior observer had a chance of shining an infinite amount
of energy on the infalling observer in what, from the point of view of
the latter, is a finite amount of time. This suggests an instability
can occur. This has been confirmed in classical general relativity
using perturbation theory and numerical analysis.

In a quantum space-time the above argument gets modified by the
reflections and backscatters that are implied by the quantization of
space-time that we discussed above. To begin with, not all light
enters the horizon to reach the infalling observer. Some is
backscattered outside the black hole towards $scri^+$. Some light crosses
the horizon and backscattering continues in the interior towards the
Cauchy horizon. At this heuristic level this is not enough to argue
that the Cauchy horizon is stabilized, but it clearly suggests that a
rethinking of the situation in a quantum space-time is in order. This
however, significantly exceeds the scope of this paper, as it would
require studying back reaction of perturbations at a quantum level,
something that is not possible in loop quantum gravity today, though
it may become feasible in a relatively near future. Since the
backscattering starts only very close to the horizon, the
backscattered light would become visible only in the remote future to
external observers, so it will not conflict with black hole
observations.

\section{Summary}

We have showed that one can complete the Dirac quantization procedure
using loop quantum gravity techniques for spherically symmetric
electrovacuum space-times. The space of physical states can be found
in closed form. Dirac observables can be identified and the physical
states labeled with their eigenvalues. The singularity is resolved due
to quantum effects as had been observed in the vacuum case. The
fundamental discreteness of space-time opens new possibilities in
analyzing the stability of the Cauchy horizon inside the
Reissner-Nordstrom black hole.

This work was supported in part by grant NSF-PHY-1305000, funds of the
Hearne Institute for Theoretical Physics, CCT-LSU and Pedeciba.


\begin{thebibliography}{9}

\bibitem{sphericalprl}
  R.~Gambini and J.~Pullin,
  Phys.\ Rev.\ Lett.\  {\bf 110}, no. 21, 211301 (2013)
  [arXiv:1302.5265 [gr-qc]]

\bibitem{bengtsson}
 I.~Bengtsson,
  Class.\ Quant.\ Grav.\  {\bf 5}, L139 (1988).

\bibitem{boswi}
  M.~Bojowald and R.~Swiderski,
  Class.\ Quant.\ Grav.\  {\bf 23}, 2129 (2006)
  [arXiv:gr-qc/0511108].


\bibitem{spherical} 
  M.~Campiglia, R.~Gambini and J.~Pullin,
  Class.\ Quant.\ Grav.\  {\bf 24}, 3649 (2007)
  [gr-qc/0703135].

\bibitem{loukowintershilt}
  J.~Louko and S.~N.~Winters-Hilt,
  Phys.\ Rev.\ D {\bf 54}, 2647 (1996)
  [gr-qc/9602003].

\bibitem{saeed}
  R.~Gambini, J.~Pullin and S.~Rastgoo,
  Class.\ Quant.\ Grav.\  {\bf 26}, 215011 (2009)
  [arXiv:0906.1774 [gr-qc]].

\bibitem{kieferlouko}
  C.~Kiefer and J.~Louko,
  Annalen Phys.\  {\bf 8}, 67 (1999)
  [gr-qc/9809005].

\bibitem{olmedo}
  R.~Gambini, J.~Olmedo and J.~Pullin,
  Class.\ Quant.\ Grav.\  {\bf 31}, 095009 (2014)
  [arXiv:1310.5996 [gr-qc]].

\bibitem{hawkingus}
  R.~Gambini and J.~Pullin,
  Class.\ Quant.\ Grav.\  {\bf 31}, 115003 (2014)
  [arXiv:1312.3595 [gr-qc]].

\bibitem{stoica}
  O.~C.~Stoica,
  Phys.\ Scripta {\bf 85}, 055004 (2012)
  [arXiv:1111.4332 [gr-qc]].

\bibitem{instability}
  E.~Poisson and W.~Israel,
  Phys.\ Rev.\ D {\bf 41}, 1796 (1990);
L. Burko and A. Ori ``Internal Structure of Black Holes and Spacetime
Singularities. An International Research Workshop, Haifa, June 29-July
3, 1997'', Institute of Physics, Bristol (UK), (1997).

\end{thebibliography}
\end{document}